\documentstyle[12pt,procsla,psfig]{article}

  \catcode`\@=11
  \long\def\@makefntext#1{
  \protect\noindent \hbox to 3.2pt {\hskip-.9pt
  $^{{\ninerm\@thefnmark}}$\hfil}#1\hfill}              

  \def\@makefnmark{\hbox to 0pt{$^{\@thefnmark}$\hss}}  

  \def\ps@myheadings{\let\@mkboth\@gobbletwo
  \def\@oddhead{\hbox{}
  \rightmark\hfil\ninerm\thepage}
\def\@oddfoot{}\def\@evenhead{\ninerm\thepage\hfil
  \leftmark\hbox{}}\def\@evenfoot{}
  \def\sectionmark##1{}\def\subsectionmark##1{}}

\begin{document}

\hskip 8.9cm FUB-HEP/96-5
\vskip 0.6cm

  \centerline{\normalsize\bf SINGLE SPIN ASYMMETRIES IN INCLUSIVE}
  \baselineskip=20pt
  \centerline{\normalsize\bf HIGH ENERGY 
   HADRON-HADRON COLLISION PROCESSES
   \footnote{talk given at the Adriatico Research Conference: 
   Trends in Collider Spin Physics, December 5-8, 1995, Trieste, Italy} }
  \baselineskip=18pt
  \centerline{\footnotesize Liang Zuo-tang}
  \baselineskip=13pt
  \centerline{\footnotesize\it Institut f\"ur theoretische Physik der 
  Freien Unbiversit\"at Berlin} 
   \baselineskip=12pt
  \centerline{\footnotesize\it  Arnimallee 14, 14195 Berlin, Germany}

  \vspace*{0.6cm}
\abstracts{
Characteristics of the 
available data are briefly summarized. 
Different theoretical approaches are reviewed with 
special attention to a non-perturbative 
model which explicitly 
takes the orbital motion of the valence quarks into account. 
The connection between such asymmetries and 
hyperon polarization in unpolarized reactions 
is discussed.} 

  \normalsize\baselineskip=15pt
  \setcounter{footnote}{0}
  \renewcommand{\thefootnote}{\alph{footnote}}
        
\section {Introduction}

Single spin hadron-hadron collision experiments, 
in which one of the colliding hadrons 
is transversely polarized, 
are of particular interests
for the following reasons: 

(A) The experiments are conceptually very simple. 

(B) The observed effects are very striking.

(C) Theoretical expectations based on pQCD 
     deviate drastically from the data. 

(D) Information on transverse spin distribution and 
    that on its flavor dependence can be obtained from 
    such experiments.

A large amount of data is now available [1-8]. 
Besides the well known 
analyzing power in $pp$-elastic scattering [1], 
we have now data [2-8] on
left-right asymmetry $A_N$ in 
single-spin inclusive processes  
for the production of different kinds of mesons, 
$\Lambda$ hyperon or direct photon in collisions 
using transversely polarized proton 
as well as antiproton beams. 
It has been observed that $A_N$ 
is up to $40\%$ in the beam fragmentation region, 
whereas the theoretical expectation [9] 
were $A_N\approx 0$.

In this talk, I will concentrate on inclusive 
hadron production and arrange the talk as follows: 
After this introduction, 
I will briefly summarize 
the characteristics of the existing data, 
the pQCD based hard scattering models,  
and the main ideas and results for $A_N$ of 
a non-perturbative approach of the Berliner group. 
They are given in section 2, 3 and 4 respectively [10].
In section 5, I will discuss the connection of $A_N$ to 
hyperon polarization in unpolarized reactions. 

\section {Characteristics of the data}

Data on $A_N$ for hadron production at high energies 
is now available for 
$p(\uparrow )+p(0)\to 
\pi $ (or $\eta$,  or $K$,  or $\Lambda ) +X$ [3-7],
$\bar p(\uparrow )+p(0)\to \pi$ (or $K) +X$ [5,7], 
and $\pi +p(\uparrow \nobreak)\to \pi$ (or $\eta ) +X$ [8]. 
These data show the following characteristics: 

(1) $A_N$ is significant in, and only in, 
   the fragmentation region of the 
   polarized colliding object and 
   for moderate transverse momenta. 

(2) $A_N$ depends on the flavor quantum number of the 
   produced hadrons. 

(3) $A_N$ depends also on the flavor 
   quantum number of the polarized projectile.

(4) $A_N\approx 0$ in the beam fragmentation
  region in $\pi ^- +p(\uparrow )\to \pi ^0 $ or $\eta +X$. 

\section{PQCD based hard scattering models}

A number of mechanisms [9-21]
have been proposed recently which can give non-zero $A_N$'s
in the framework of QCD and
quark or quark-parton models.
They can approximately be divided
into two categories:
(1) perturbative QCD based hard scattering models [9,11-15]
and (2) non-perturbative quark-fusion models [16-21].

In the pQCD based hard scattering models [9,11-15],
the cross section for inclusive hadron production
in hadron-hadron collision 
is expressed as convolution of the following three factors:
the momentum distribution functions of the constituents 
(quarks, antiquarks or gluons)
in the colliding hadrons;
the cross section for the elementary hard scattering
between a constituent of one of the colliding hadrons
with one of the other; and
the fragmentation function of the scattered constituent. 
The cross section for the elementary hard scattering
can be and has been calculated [9] using pQCD.
The obtained result [9] shows that, to the leading order,
the asymmetry for the elementary process
is proportional to $m_q/\sqrt{s}$
(where $m_q$ is the quark mass and $\sqrt{s}$ is the
total c.m. energy of the colliding hadron system),
which is negligibly small at high energies.
Hence, to describe
the observed $A_N$'s in terms of
such models, one can make use one or more
of the following three possibilities:
(i) Look for higher order and/or higher twist
 effects in the elementary processes
 which lead to larger asymmetries;
(ii) Introduce asymmetric
  intrinsic transverse momentum distributions
  for the transversely polarized quarks
  in a transversely polarized nucleon;
(iii) Introduce asymmetric transverse momentum distributions
  in the fragmentation functions for the transversely polarized quarks. 
  All three possibilities have been discussed in the literature [11-15].
  We note that the question whether (or which one of)
  these possible effects indeed exist(s) is not yet settled.
  It is clear that
  under the condition that pQCD is indeed applicable for
  the description of such processes,
  it should (at least in principle) be possible to
  find out how significantly the effects mentioned in (i)
  contribute to $A_N$  by performing
  the necessary calculations.
But, in contrast to this,
the asymmetric momentum distributions 
mentioned in (ii) and (iii)
have to be introduced by hand. 
Whether such asymmetries indeed exist,
and how large they are if they exist,
are questions which
can only be answered by
performing suitable experiments [22].
     
\section {A non-perturbative phenomenological approach} 

In this section, I discuss a non-perturbative 
phenomenological approach.  
This section is based on 
the work done with C. Boros and T. Meng [16-21].

\subsection{$A_N$ for $\pi$ production}

To understand the observed asymmetries, 
we took a close look at the data and 
noted the following.

First, $A_N$ is 
significant only in the 
fragmentation region. 
This is a region where 
soft processes dominate. 
It is therefore clear that 
non-perturbative effects should contribute significantly;  
and it is not surprising to see that 
the expectations based on pQCD 
deviate so drastically from the data. 

Second, the fact that $A_N$ has been observed mainly in 
the fragmentation region, that it depends on 
the flavor quantum numbers of the produced
hadrons and on that of the projectile explicitly   
shows that the valence quarks play the dominating role. 
We recall that valence quarks should be treated as Dirac particles 
and orbital angular momentum is {\it not} 
a good quantum number in Dirac theory.   
Hence, orbital motion is always involved for  
the valence quarks even when they are in their ground states. 
The direction of the orbital motion is 
determined uniquely by the polarization of the quarks. 
The polarization of the valence quarks is determined 
by the the wave function of the nucleon. 
This implies that, for proton, 
$5/3$ of the 2 $u$ valence quarks are polarized in the same,  
and $1/3$ in the opposite, direction as the proton. 
For $d$, they are $1/3$ and $2/3$ respectively.

Third, from unpolarized experiments 
we learned that mesons in the fragmentation region 
are predominately produced 
through the direct formation (fusion) of 
the valence quarks of the projectile $q_v^P$ with 
suitable antiseaquarks $\bar q_s^T$ associated with 
the target. (See [17-21] and the references cited therein 
for detailed discussions).

Fourth, since hadron are extended objects, 
geometrical effects should play a significant role 
also in single-spin processes. 
We expect a significant surface effect, 
which implies that only the mesons directly 
formed near the front surface 
of the projectile keep the information of polarization.

Having these four points in mind, we obtain 
a correlation between 
the polarization of the valence quarks 
and the direction of transverse motion 
of the produced mesons. More precisely, we obtain that
mesons produced through the direct formation 
of upwards transversely polarized valence quarks of the projectile 
with suitable antiseaquarks associated with the target 
have large probability to go left. 
Hence, once we know the polarization of the projectile, 
we can use the baryon wave function 
to determine the polarization 
of its valence quarks and 
then $A_N$ for the produced hadron. 
E.g., for $p(\uparrow )+p(0)\to \pi +X$, 
we obtain: 
\begin{itemize}
\itemsep=-0.10truecm
\item \vskip -0.2truecm $A_N[\ p(\uparrow )+p(0)\to \pi ^++X] >0,$ 
\ \ \ $A_N[\ p(\uparrow )+p(0)\to \pi ^-+X] <0,$ 
\item $0<A_N[\ p(\uparrow )+p(0)\to \pi ^0+X] <A_N[\ p(\uparrow )+p(0)\to \pi ^++X] $,
\item The magnitudes of all these $A_N$'s increase with increasing
$x_F$. (Here $x_F\equiv 2p_\|/\sqrt{s}$, $p_\|$ is the longitudinal 
momentun of the pion.)
\end{itemize}\nopagebreak
\vskip -0.12truecm \noindent
{\it All these qualitative features are 
in good agreement with the data!}

Quantitatively, $A_N$ can be expressed [18] in terms of
the spin-dependent quark distributions $q_v^\pm(x|s,tr)$ 
in transversely polarized nucleon and other quantities which 
can be measured in unpolarized experiments. 
A rough estimation of 
$A_N$ was made by assuming $q_v^\pm (x|s,tr)\propto q_v(x|s)$.  
The result is shown in Fig.1.

\begin{tabular}{ll}
\begin{minipage}{6cm}
Fig.1: $A_N$ as a function of $x_F$
for $p(\uparrow )+p(0)\to \pi +X$, $p(\uparrow )+p(0)\to l\bar l+X$
and $\bar p(\uparrow )+p(0)\to l\bar l +X$ at 200 GeV/c.
The data for pions are taken from [4].
For lepton pairs, the solid and dash-dotted curves correspond
to $Q=4$ and 9 GeV/c, respectively.
\end{minipage}
 &
\begin{minipage}{8cm}
\psfig{file=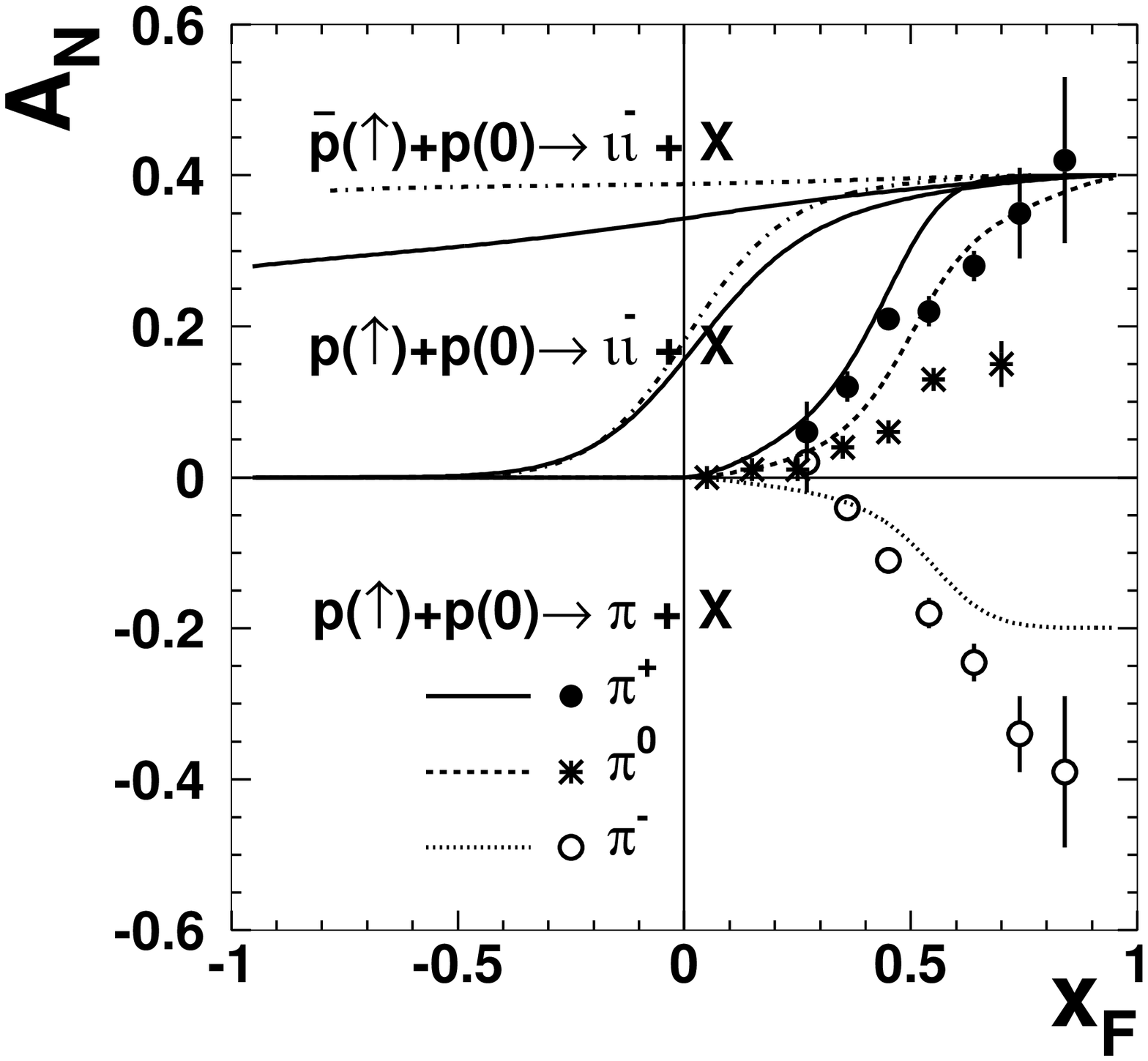,height=8cm}
\end{minipage}
\end{tabular}
\vskip -0.6cm 
\noindent 
It has also been found that 
there exist many simple relations 
between the $A_N$'s for hadron production 
in reactions using different projectile-target combinations.
E.g., comparing $\bar p(\uparrow )+p(0)$ 
with $p(\uparrow )+p(0)$, we predicted [16] that 
there should be change of sign for $\pi ^+$ and $\pi^-$ 
productions, which is confirmed by the E704 data [5,7]. 
Predictions for processes 
using polarized neutron or deutron targets 
have also been made [18]. 
They can be tested by future experiments.

\vskip -0.1cm
\subsection{$A_N$ for $K$ production}

\vskip -0.1cm
The above-mentioned analysis can be extended to $K$-mesons 
in a straightforward manner. The results are summarized in [18,21]. 
Together with $K_s^0\approx {1 \over \sqrt{2}}(K^0+\bar K^0)$, 
the results there imply in particular that 
$A_N[\ \bar p(\uparrow )+p(0)\to K^0_s+X] \approx 
A_N[\ p(\uparrow \nobreak)+p(0\nobreak)\to K^0_s+X],$   
and they should be negative. 
This prediction in [18] has also been confirmed by the 
E704 data [7].

\vskip -0.1cm
\subsection{$A_N$ for lepton-pair production}

\vskip -0.1cm
Not only mesons but also 
lepton pairs should exhibit left-right asymmetry 
in single-spin processes. 
Here, since the production mechanism 
--- the well known Drell-Yan mechanism --- 
is very clear, 
and there is no fragmentation, 
measurement of such asymmetry provide a crucial test 
of the model and is also very helpful to 
distinguish the origin of the observed $A_N$. 

The calculation for $A_N$ for lepton-pair 
is straightforward [18]. 
Here I just include the 
results for $p(\uparrow )+p(0)$ 
and $\bar p(\uparrow )+p(0)$ collisions 
at 200 GeV/c in Fig.1.

\vskip -0.1cm
\subsection{$A_N$ for hyperon production}

\vskip -0.1cm
$A_N$ for $\Lambda$ production [6,7]  
is of particular interest for the following reason: 
$\Lambda$ in the large $x_F$ region 
comes predominately from the 
hadronization of a spin-zero diquark. 
The fragmentation function has to be symmetric w.r.t. the 
moving direction of this diquark. 
The existence of $A_N$ can only be understood 
in terms of other effects, in particular, the 
picture described above.
Here, we have the following three possibilities 
for direct formations: 
(a) $(u_vd_v)^P+s_s^T\to \Lambda $;  
(b) $u_v^P+(d_ss_s)^T\to \Lambda $; and 
(c) $d_v^P+(u_ss_s)^T\to \Lambda $;  
where $v$ stands for valence, $s$ for sea, 
$P$ and $T$ for projectile and target respectively.
According to the proposed picture, 
$\Lambda $'s produced from (b) 
should have large probabilities to go left 
and thus give positive contributions to $A_N$, 
while those from $(c)$ contribute negatively to it. 
(a) should be 
associated with the production of 
a meson directly formed through fusion of 
the $u$ valence quark of the projectile 
with an anti-sea-quark of the target. 
This meson should 
have a large probability to 
obtain an extra transverse momentum to the left.  
Thus, due to momentum conservation, 
the $\Lambda$ from (a) should 
have a large probability to 
obtain an extra transverse momentum to the right.
This implies that $(a)$ contributes negatively to $A_N$, 
opposite to that of the associatively produced meson 
($\pi ^+$ or $K^+$ or other).
It has been shown that 
(a) plays the dominating role in the large, 
(b) and (c) in the middle and non-direct formation in the 
small, $x_F$ region. 
The interplay of these different contributions  
leads exactly to the observed $x_F$ dependence of $A_N$. 
A quantitative estimation has been made and is shown ion Fig.2. 
Similar analysis have also been made for 
other hyperons. The results can be found in [20]. 

\begin{tabular}{ll}
\begin{minipage}{5cm}
Fig.2:
$A_N$ as a function of $x_F$ 
for $p(\uparrow)+p(0)\to \Lambda +X$ at $200$ GeV/c.
The data are taken from [6,7]. 
\end{minipage}
&
\begin{minipage}{8cm}
\psfig{file=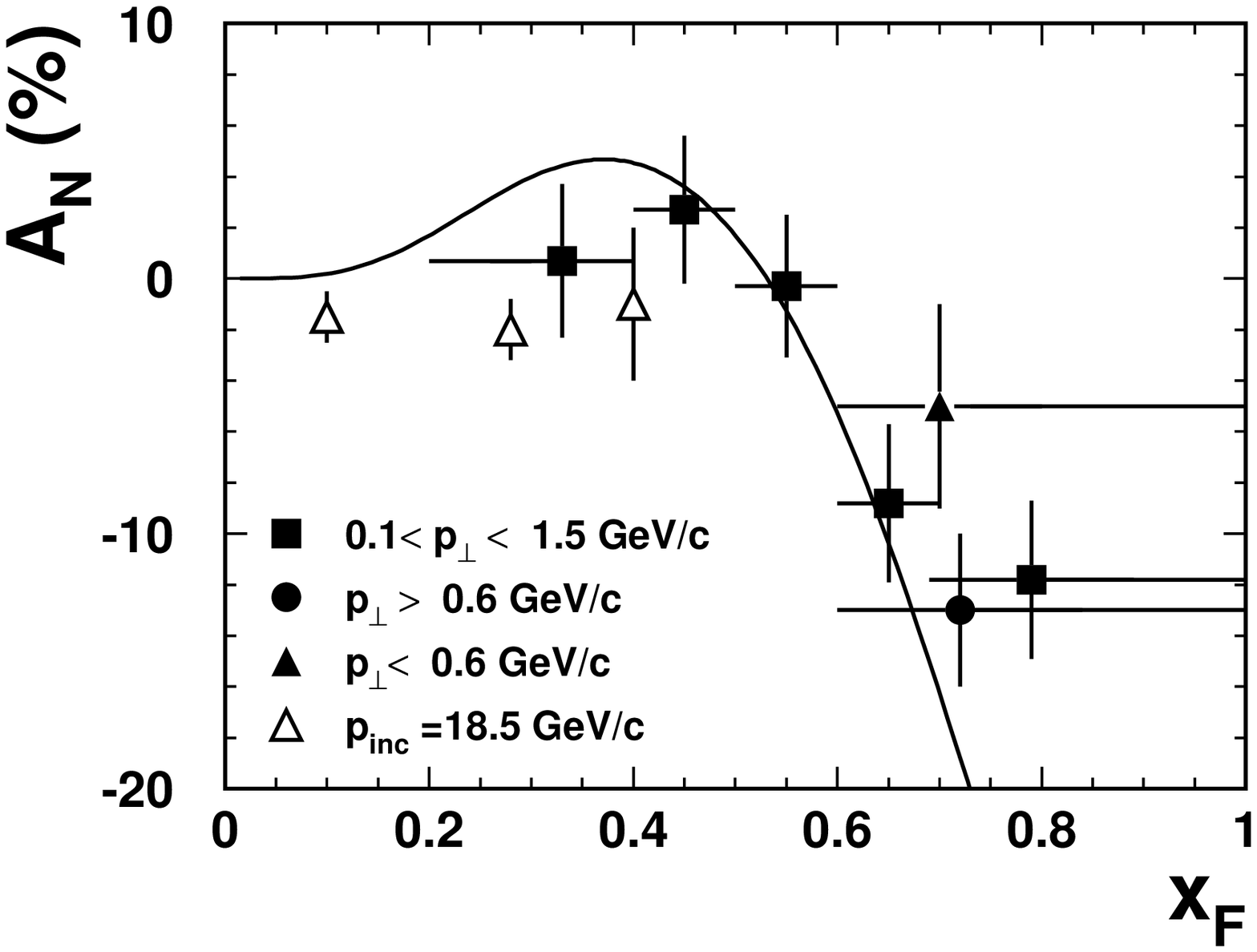,height=6cm}
\end{minipage}
\end{tabular}

Before I end this section, I 
would like to emphasize that 
the asymmetries are expected 
to have the following in common:
($\alpha $) $A_N$ for hadron  
is expected to be significant only 
in the fragmentation region of the polarized colliding object.
It should be zero in the fragmentation region of the 
unpolarized one. 
($\beta $) $A_N$ for hadron 
in the fragmentation regions of the polarized colliding objects 
depends little on what kinds of unpolarized 
colliding objects are used.
($\gamma$) In contrast, $A_N$ for lepton-pair production 
depends not only on the polarized colliding object but also 
on the unpolarized one. 
Point($\alpha$) implies in particular that $A_N=0$ 
in the beam fragmentation region in 
$\pi ^-+p(\uparrow )\to \pi^0 $ or $\eta +X$. 
This is predicted in [16-18]
and has been confirmed by the experiment [8].

\section{Hyperon polarization in unpolarized hadron-hadron collisions}

Another kind of striking spin effect in 
inclusive hadron production at high energies 
is hyperon polarization $P_H$. 
It has been observed that hyperons produced 
in high hadron-hadron or hadron-nucleus collisions
are polarized 
transversely to the production plane 
although neither the projectile nor 
the target were polarized before the collisions.  
Data are available for the production of 
different kinds of hyperons 
in $pp$ and $p$-nucleus collisions [23] and also  
in $K^-$-nucleus [24] or $\Sigma^-$-nucleus collisions [25]. 
They show the following characteristics: 

(1) $P_H$ is significant in and only in the fragmentation region.

(2) $P_H$ depends on the flavor quantum number of the produced hyperon.

(3) $P_H$ depends on the flavor quantum number of the projectile. 

(4) $P_H$ in the beam fragmentation 
    region depends little on the targets.

Not only the similarities between 
these characteristics of the data with those for $A_N$ 
but also the following seem to suggest that 
these two kinds of phenomena 
are closely related to each other.
First of all, we note: 
$A_N\not=0$ means that there is a 
correlation between the direction of transverse 
motion of the produced hadron and 
the polarization of the projectile. 
$P_H\not=0$ shows that there is a
correlation between the direction of transverse 
motion of the produced hyperon and 
the polarization of this hyperon. 
Both of them describe the correlation between 
transverse motion and transverse polarization.  
It can easily be impgined that 
the polarization of the hyperons observed in 
the fragmentation region  
should be closely related to that of the projectile. 
(This is true even for $\Lambda$, a fact observed 
by E704 collaboration recently [7].)
Hence, there should also be a close relation between 
$A_N$ and $P_H$. 
Second, crossing symmetry tell us the following: 
If hadron $C$ in reaction $A(\uparrow)+B(0)\to C+X$ has 
large probability to go left, 
$\bar A$ in reaction $\bar C(0)+B(0)\to \bar A+X$
should be polarized upwards.

In the picture we have discussed in last section, 
there is a relation between the transverse moving direction 
of the produced hadrons and the polarizations of the valence quarks.
More precisely, we have: 
(I) Hadron produced through 
$q_v(\uparrow\nobreak )^P+\bar q_s^T [\mbox {or\ } (q_sq_s)^T ]\to M$ [or $B$]
has large probability to go left. 
(II) Baryon produced through 
$(q_vq_v)^P+q_s^T \to B$ is 
associated with $(q_v^a)^P+\bar q_s\to M$ and 
has large probability to move in the opposite direction as $M$ does.
We have seen that 
these relations describe $A_N$ very well. 
Now we show [26] that they describe also $P_H$.

We take $\Lambda $ as an example.
As described in section 4.4, for large $x_F$, 
the direct formation process (a), $(u_vd_v)^P+s_s^T\to \Lambda$, dominate. 
This process is associated with $(u_v^a)^P+\bar s_s^T\to K^+$. 
If $\Lambda$ is going left, 
the associatively produced $K^+$ 
should have large probability to go right. 
According to (I), 
this implies that $(u_v^a)^P$ has large
probability to be downwards polarized. 
Since $K$ has spin zero, 
$\bar s_s^T$ should be upwards polarized. 
Thus the $s_s^T$ should downwards polarized 
if sea is not polarized.  
This implies that $\Lambda $ should have large
probability to be downwards polarized, i.e. $P_\Lambda <0$. 
Taking the contributions from other kinds of direct formation  
and non-direct-formation processes into account, 
we calculated [26] $P_\Lambda $ as a function of $x_F$ 
without any new parameters. 
The result is shown is Fig.3. 

\begin{tabular}{ll}
\begin{minipage}{5cm}
Fig.3: Lambda polarization 
$P_\Lambda $ as a function of $x_F$.
The data are takne from [23] 
or the papers cited therein. 
\end{minipage}
&
\begin{minipage}{8cm}
\psfig{file=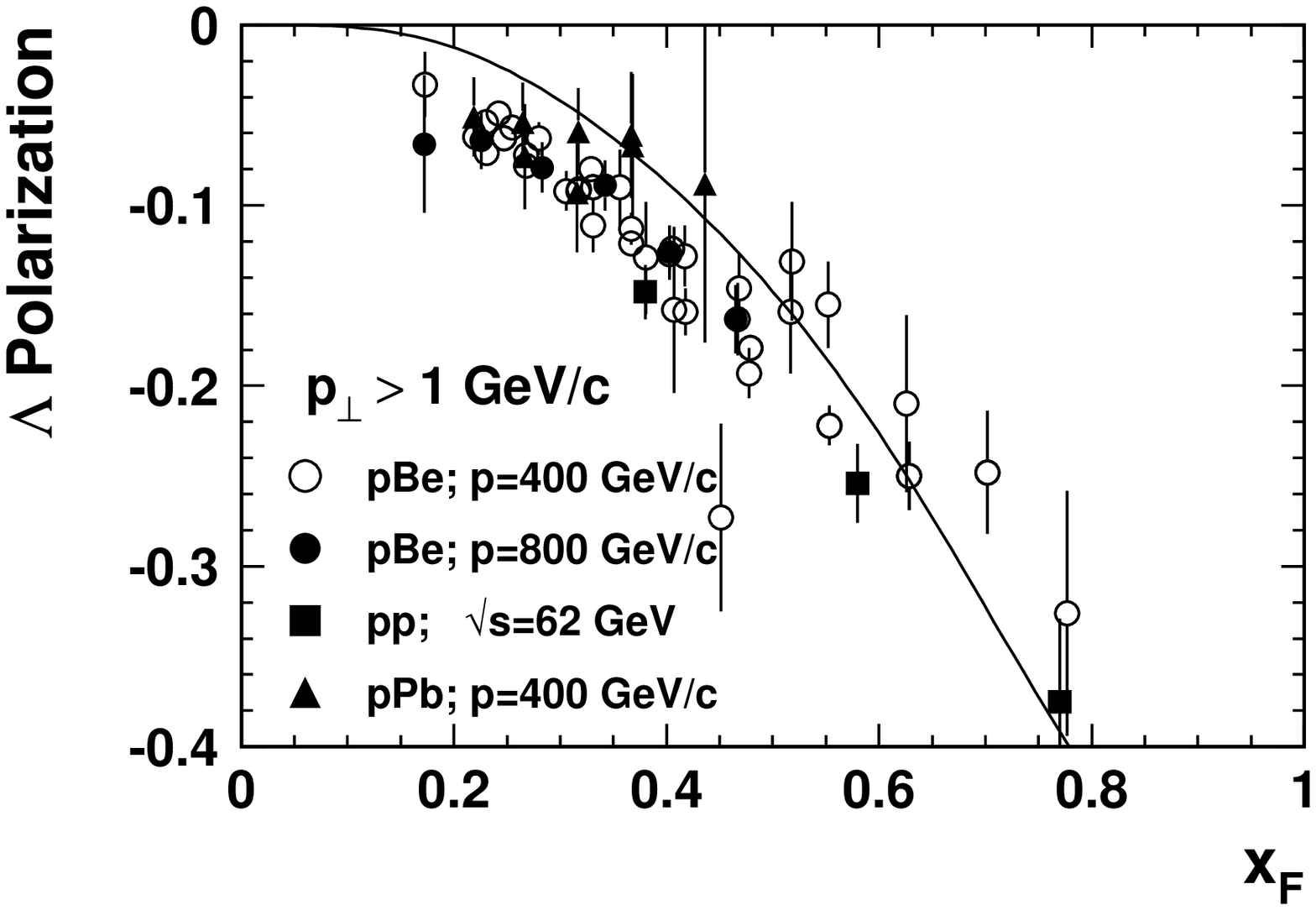,height=6cm}
\end{minipage}
\end{tabular}

Similar analysis can be made for other hyperons. 
All the qualitative results are consistent with the
available data. 
We have also considered the processes using 
$K^-$- or $\Sigma ^-$-beams, and we obtained that 
$P_\Lambda$ in $K^-+A\to \Lambda +X$ 
should be large and positive whereas 
that in $\Sigma^-+A\to \Lambda +X$ 
should be small and negative. 
These are also consistent with the 
recent experimental observations [24,25]. 
 
It should also be mentioned that,  
with this picture in mind, 
it is not surprising to see that there 
is [7] a correlation between 
the polarization of the projectile 
and the produced $\Lambda$. 
Although the $ud$-diquark is in 
a spin-zero state thus does not 
carry any information about polarization, 
the left $u$-valence quark in proton determines 
the polarization of the proton.  
Because of the mechanism of 
associated production described above, 
the polarization of this $u$ valence quarks 
determines the polarization 
of the $s_s$ quark which combine with the 
$ud$-diquark to form the $\Lambda$. 
Thus we get a close relation between the polarization of 
the proton and the produced $\Lambda$.
This is consistent with the E704 data [7].

Last but not least, I would like to mention that 
the existence of these striking $A_N$ and $P_H$ 
in hadron-hadron collisions 
can be used to study the hadronic behavior of 
particles in other processes.  
It has been suggested [27] that they can be 
used as sensors to test whether 
the virtual photon in deep inelastic 
scattering behaves like a hadron. 
Details can be found in [27].


\vskip 0.2truecm

\begin{thebibliography}{9}
\bibitem{ref} See, e.g., 
    P.R. Cameron et al., Phys. Rev. {\bf D32},3070 (1985);
    A.D. Krisch, in {\it High-Energy Spin Phys.}, 
    Proc. of the 9th Inter. Symp., 1990,
    Bonn, ed. K.H. Althoff and W. Meyer, Springer Verlag (1991), p.20;
    and the references cited therein.
\bibitem{ref}  S.~Saroff et al., Phys. Rev. Lett. {\bf 64},995 (1990). 
\bibitem{ref}  FNAL E581/704 Collab.,
    D.L. Adams et al., Phys. Lett. {\bf B261}, 201 (1991).  
\bibitem{ref}  FNAL E704 Collab.,
      D.L.~Adams et al., Phys. Lett. {\bf B264}, 462 (1991);
      {\bf B276}, 531 (1992);
      Z. Phys. {\bf C56},181 (1992). 
\bibitem{ref} A. Yokosawa, in
      {\it Frontiers of High Energy Spin Physics},
     Proc. of the 10th Inter. Symp., Nagoya, Japan 1992,
     edited by T. Hasegawa {\it et al.}.  
\bibitem{ref} FNAL E704 Collab., 
    A.Bravar et al., Phys. Rev. Lett. {\bf 75}, 3073 (1995).  
\bibitem{ref} A. Bravar, these Proceedings; and the references given there.
\bibitem{ref} V.D. Apokin et al., Phys. Lett. {\bf B243}, 461 (1990); and
  ``$x_F$-dependence of the asymmetry in inclusive $\pi ^0$ and
  $\eta ^0$ production in beam fragmentation region",
  Serpuhkov-Preprint (1991).
\bibitem{ref}  G.~Kane, J.~Pumplin and W.~Repko,
   Phys. Rev. Lett. {\bf 41}, 1689 (1978).
\bibitem{ref}  A review of these aspects can also be found in, 
         Meng Ta-chung,  in Proc. of
         the Workshop on the Prospects of
         Spin Physics at HERA, August 28-31, 1995, 
         DESY Zeuthen. edited by 
         J. Bl\"umlein and W.D. Nowak, p.121.
\bibitem{ref}  D. Sivers, Phys. Rev. {\bf D41},83 (1990);
                 {\bf D41}, 261 (1991).
\bibitem{ref}  J.~Qiu and G.~Sterman, Phys. Rev. Lett. {\bf 67}, 2264 (1991).
\bibitem{ref}  J.~Collins, Nucl. Phys. {\bf 396}, 161 (1993); 
               J.~Collins, S.F. Hepplelmann and G.A. Ladinsky,
                  Nucl. Phys. {\bf B420}, 565 (1994).
\bibitem{ref}  A.Efremov, V.Korotkiyan, O.Teryaev, 
               Phys. Lett. {\bf B 348}, 577 (1995).
\bibitem{ref} M.~Anselmino; these proceedings; and the references given there.
\bibitem{ref} Meng Ta-chung, in Proc. of the 4th Workshop on
     High Energy Spin Physics,
    Protvino, Russia, Sept. 1991, ed. S.B. Nurushev, p. 121 (1991).
\bibitem{ref}  Liang Zuo-tang and Meng Ta-chung, 
      Z. Phys. {\bf A344}, 171 (1992). 
\bibitem{ref}  C.~Boros, Liang Zuo-tang and Meng Ta-chung,
               Phys. Rev. Lett. {\bf 70}, 1751 (1993); 
               Phys. Rev. {\bf D51},4698 (1995).
\bibitem{ref}  Liang Zuo-tang and Meng Ta-chung, 
               Phys. Rev. {\bf D49},3759 (1994).
\bibitem{ref}  C. Boros and Liang Zuo-tang, Phys. Rev. {\bf D53}, R2279 (1996).
\bibitem{ref}  C.~Boros, Liang Zuo-tang and Meng Ta-chung, FUB-HEP/96-1.
\bibitem{ref}  C.~Boros, Liang Zuo-tang, Meng Ta-chung and R. Rittel, 
               FUB-HEP/96-4.
\bibitem{ref}  For a review of the data before 1990,  
      see, e.g., K. Heller, in {\it High Energy Spin Physics}, 
      Proc. of the 9th Inter. Symp., Bonn, Germany,
      1990, edited by K.H. Althoff, W. Meyer (Springer-Verlag, 1991); 
      More recent data can be found, e.g., in  
      FNAL E761 Collab., A. Morelos et al.,
      Phys. Rev. D{\bf 52}, 3777 (1995); and the references given there.  
\bibitem{ref}  S.A. Goulay et al., Phys. Rev. Lett. {\bf 56}, 2244 (1986); 
      and the references given there.
\bibitem{ref}  CERN WA89 Collab., M.I. Adamovich et al., 
          Z. Phys. {\bf A350}, 379 (1995). 
\bibitem{ref} C. Boros, Liang Zuo-tang and Meng Ta-chung, in
          preparation. 
\bibitem{ref} C. Boros, Liang Zuo-tang and Meng Ta-chung, FUB-HEP/95-21.
\end{thebibliography}
\end{document}